\documentclass[times,10pt,twocolumn]{article}
\usepackage{latex8}
\usepackage{times}

\usepackage{amsmath,amssymb}
\usepackage{graphicx}
\usepackage{amsthm}

\newfont{\bg}{cmr9 scaled\magstep4}
\newtheorem{theorem}{Theorem}

\theoremstyle{remark}
\newtheorem{example}{Example}

\title{On a Geometric Structure of Pure Multi-qubit Quantum States and
Its Applicability to a Numerical Computation}

\author{Kimikazu Kato
\\{\it Nihon Unisys, Ltd.}
\\{\it \&}
\\{\it Department of Computer Science,}
\\{\it University of Tokyo}\\{\it kkato at is.s.u-tokyo.ac.jp} 
\and \ Mayumi Oto
\\{\it Toshiba Corporation}
\\{\it mayumi.ooto at toshiba.co.jp}
\and \ Hiroshi Imai
\\{\it Department of Computer Science,}
\\{\it University of Tokyo}
\\{\it \&}
\\{\it ERATO-SORST Quantum}
\\{\it Computation and Information}
\\{\it imai at is.s.u-tokyo.ac.jp} 
\and \ Keiko Imai
\\{\it Department of Information and} 
\\{\it System Engineering, Chuo University}
\\{\it imai at ise.chuo-u.ac.jp}
}

\date{}

\def\Tr{{\rm Tr}\;}

\begin{document}
\maketitle
\pagestyle{empty}
\thispagestyle{empty}

\begin{abstract}
For one-qubit pure quantum states, it is already proved that the
Voronoi diagrams with respect to two distances --- Euclidean distance
and the quantum divergence --- coincide. This fact is a support for a known
method to calculate the Holevo capacity. 
To consider an applicability of this method to quantum states of a
 higher level system, it is essential to check if the coincidence of the
 Voronoi diagrams also occurs. In this paper, we show a negative result
 for that expectation. In other words, we mathematically prove that
 those diagrams no longer coincide in a higher dimension. That indicates
 that the method used in one-qubit case to calculate the Holevo capacity
 might not be effective in a higher dimension.
\end{abstract}

\Section{Introduction}

The movement of trying to apply quantum mechanics to information
processing has given vast research fields in computer
science \cite{NC00}. Especially among them quantum information theory is
developed as one of the richest research fields. Some aspect of quantum
information theory is to investigate a kind of distance between two
different quantum states. Depending on the situation, several distances
are defined in quantum states.

The significance of quantum information theory is also due to its rich
variety of mathematical methods used there. Especially, geometric
interpretation of quantum states has been an important theme. The
researches from that kind of view are categorized as ``quantum information
geometry.'' Actually some properties of a space of quantum states as a
metric space have been researched in some
contexts \cite{kato05,MW01,petz96}.

A quantum channel is a channel that transfers quantum
information. Mathematically it is represented as an affine
transformation between two Hilbert spaces, each of which is a
representation of quantum states. One of the main problem in the quantum
information theory is how well a quantum channel transfers
information. Especially, one of the most important indications that show
a capacity of quantum channel is the Holevo
capacity \cite{holevo98}. Intuitively speaking, the Holevo capacity indicates
how much the channel preserves the size of the space of the quantum
states. The measure used here is called the quantum divergence, which is
a kind of a distance of two quantum states.

Oto et al.~\cite{oto04} showed that the Holevo capacity of one-qubit
states can be numerically computed by considering the image of the
source points of the channel. In that paper, points in pure states are
plotted so that they are almost uniformly distributed with respect to
the Euclidean distance. Here pure states are important because they
appear in the boundary of the convex object which corresponds to the
whole space of the quantum states. Although their images are
dealt in a context of the divergence, the algorithm is reasonable
because the spaces of pure quantum states with respect to the two
distances have the same structure. In other words, in pure states,
uniformly distributed points in the world of the Euclidean distance is
also uniformly distributed with respect to the divergence.  The authors
showed that fact mathematically considering Voronoi
diagrams \cite{kato05}.

A natural question that arises after this story is ``What happens in a
higher level systems?'' If you could say the same thing in a higher
dimension, the method used in the one-qubit case for a calculation of
the Holevo capacity might be applied to a general case. However,
unfortunately we found it is not the case. The main result of this
paper is the fact that in a higher dimension, the two distance spaces
--- the space of pure states with respect to the Euclidean distance and the
one with respect to the divergence --- no longer have the same
structure. More correctly, we prove that the Voronoi diagrams
with respect to the two distances do not coincide in a higher
dimension. Additionally we give some examples for the understanding of
the structure.

The rest of this paper is organized as follows. First in Section
\ref{sec-preliminaries}, we give some basic facts in quantum information
theory. In Section \ref{sec-one-qubit}, we explain briefly the known
fact about one-qubit quantum space. Section \ref{sec-higher-dim} is the
main part of this paper, where we show some workout for a higher level
case and give some illustrative examples. Lastly in Section
\ref{sec-conclusion}, we summarize the result and give it some
discussion.

\Section{Preliminaries}\label{sec-preliminaries}
\SubSection{Parameterization of quantum states}
In quantum information theory, a density matrix is representation of
some probabilistic distribution of states of particles. A density matrix
is expressed as a complex matrix which satisfies three conditions: a) It
is Hermitian, b) the trace of it is one, and c) it must be semi-positive
definite. We denote by $\mathcal{S}(\mathbb{C}^d)$ the space of all
density matrices of size $d\times d$. It is called ``d-level system.''

Especially in two-level system, which is often called ``one-qubit
system'', the conditions above are equivalently expressed as 
\begin{eqnarray}\label{one-qubit-rho}
\lefteqn{ \rho= \left(
\begin{array}{cc}
 \displaystyle \frac{1+z}{2}& \displaystyle \frac{x-iy}{2} \\
 \displaystyle \frac{x+iy}{2}& \displaystyle \frac{1-z}{2}
\end{array}
\right),}\nonumber\\
&x^2+y^2+z^2 \leq 1,\quad x,y,z\in \mathbb{R}.
\end{eqnarray}
The parameterized matrix correspond to the conditions a) and b), and the
inequality correspond to the condition c).

There have been some attempt to extend this Bloch ball expression to a
higher level system. A matrix which satisfies only first two condition,
Hermitianness and unity of its trace, is expressed as:
{\setlength{\arraycolsep}{0mm}
\begin{eqnarray} \label{multi-level-rho}
\lefteqn{\rho=}\nonumber\\ 
&{\footnotesize
\left(
\begin{array}{ccccc}
\displaystyle\frac{\xi_1+1}{d} &
\!\!\!\displaystyle\frac{\xi_d-i\xi_{d+1}}{2} &
\cdots & &   
\displaystyle\frac{\xi_{3d-4}-i\xi_{3d-3}}{2} 
\smallskip\\
\displaystyle\frac{\xi_d+i\xi_{d+1}}{2} & 
\!\!\!\displaystyle\frac{\xi_2+1}{d} &
\cdots & & 
\displaystyle\frac{\xi_{5d-8}-i\xi_{5d-7}}{2} 
\smallskip\\
\vdots & & \!\!\!\!\ddots & & \vdots \\
\displaystyle\frac{\xi_{3d-6}+i\xi_{3d-5}}{2} & \cdots& &
\!\!\!\!\!\!\!\displaystyle\frac{\xi_{d-1}+1}{d}& 
\displaystyle\frac{\xi_{d^2-2}-i\xi_{d^2-1}}{2}\smallskip\\
\displaystyle\frac{\xi_{3d-4}+i\xi_{3d-3}}{2} & \cdots& &
\!\!\!\!\!\!\!\displaystyle\frac{\xi_{d^2-2}+i\xi_{d^2-1}}{2} &
\displaystyle
 \frac{-\sum_{i=1}^{d-1}\xi_i+1}{d}
\end{array}
\right)},\nonumber\\
&\quad \xi_i\in \mathbb{R}.
\end{eqnarray}
}
Actually, any matrix which is Hermitian and whose trace is one is
expressed this way with some adequate $\{\xi_i\}$.  This condition
doesn't contain a consideration for a semi-positivity. To add the
condition for a semi-positivity, it is not simple as in one-qubit case,
and we have to consider complicated
inequalities \cite{byrd03,kimura03}. Note that this is not the only way
to parameterize all the density matrices, but it is reasonably natural
way because it is natural extension of one-qubit case and has a special
symmetry.

Additionally our interest is a pure state. A pure state is expressed by
a density matrix whose rank is one. A density matrix which is not pure
is called a mixed state. A pure state has a special meaning in quantum
information theory and also has a geometrically special meaning because
it is on the boundary of the convex object. In one-qubit case, the
condition for $\rho$ to be pure is
\begin{eqnarray}
 x^2+y^2+z^2=1.
\end{eqnarray}
This is a surface of a Bloch ball. On the other hand, in general case,
the condition for pureness is again expressed by complicated
inequalities.  

\SubSection{The quantum divergence and the Holevo capacity}\label{holevo}
We define the log of density matrix. When eigenvalues of $\rho$
 are diagonalized with a unitary matrix $X$ as
\begin{eqnarray}
 \rho= X
\left(
\begin{array}{cccc}
 \lambda_1& & & \\
&\lambda_2&&\\
&&\ddots&\\
&&&\lambda_d
\end{array}
\right)
X^*,
\end{eqnarray}
the log of $\rho$ is defined as
\begin{eqnarray}
 \log\rho= X 
\left(
\begin{array}{cccc}
 \log\lambda_1& & & \\
&\log\lambda_2&&\\
&&\ddots&\\
&&&\log\lambda_d
\end{array}
\right)
X^*. 
\end{eqnarray}

The quantum divergence is one of measures that show the
difference of two quantum states. The quantum divergence of the two
states $\sigma$ and $\rho$ is defined as
\begin{eqnarray}
 D(\sigma||\rho) = \Tr \sigma (\log \sigma - \log \rho) .
\end{eqnarray}
Note that though this has some distance-like properties, it is not commutative, i.e.~$D(\sigma||\rho)\neq
D(\rho||\sigma)$. The divergence $D(\sigma||\rho)$ is not defined when
$\rho$ does not has a full rank, while $\sigma$ can be non-full
rank. This is because for a non-full rank matrix, a log of zero appears in
the definition of the divergence. However, since $0\log 0$ is naturally
defined as $0$, some eigenvalues of $\sigma$ can be zero.

A quantum channel is the linear transform that maps quantum states to
quantum states. In other words, a linear transform
$\Gamma:M(\mathbb{C};d)\to M(\mathbb{C};d)$ is a quantum channel if
$\Gamma(\mathcal{S}(\mathbb{C}^d))\subset \mathcal{S}(\mathbb{C}^d)$.

The Holevo capacity \cite{holevo98} of this quantum channel is known to
be equal to the maximum divergence from the center to a given point and
the radius of the smallest enclosing ball. The Holevo capacity
$C(\Gamma)$ of a 1-qubit quantum channel $\Gamma$ is defined as
\begin{eqnarray}
  C(\Gamma)= \inf_{\sigma\in \mathcal{S}(\mathbb{C}^d)} 
\sup_{\rho\in \mathcal{S}(\mathbb{C}^d)} D(\Gamma(\sigma)||\Gamma(\rho)).
\end{eqnarray}

\Section{One-qubit case}\label{sec-one-qubit}
Our first motivation to investigate a Voronoi diagram in quantum states
is the numerical calculation of the Holevo capacity for one-qubit quantum
states \cite{oto04}. In that paper, some points are plotted in the source
of channel, and it is assumed that just thinking of the images of
plotted points is enough for approximation. Actually, the Holevo capacity is
reasonably approximated taking the smallest enclosing ball of the images
of the points.  More precisely, the procedure for the approximation is
the following:
\begin{enumerate}
 \item Plot equally distributed points on the Bloch ball which is the
       source of the channel in problem.
 \item Map all the plotted points by the channel.
 \item Compute the smallest enclosing ball of the image with respect to
       the divergence. Its radius is the Holevo capacity.
\end{enumerate}
In this procedure, step 3 uses a farthest Voronoi diagram. That is the
essential part to make this algorithm effective because Voronoi diagram
is the known fastest tool to seek a center of a smallest enclosing ball of points.

However, when you think about the effectiveness of this algorithm, there might
arise a question about its reasonableness. Since the Euclidean distance and the
divergence are completely different, Euclideanly uniform points are
not necessarily uniform with respect to the divergence. So why does this
mechanism work correctly? You cannot say the approximation is good
enough unless uniformness of the image of points is guaranteed. That concern
is overcome by comparing the Voronoi diagrams. The following theorem is more
precise description \cite{kato05}.

\begin{theorem}
Suppose that $n$ one-qubit pure states are given, the following Voronoi
diagrams of them are equivalent:
\begin{enumerate}
\item
The Voronoi diagram in pure states obtained by taking a limit of the
     diagram with respect to the divergence
\item
The Voronoi diagram on the sphere with respect to the ordinary
geodetic distance
\item
The section of the three-dimensional Euclidean Voronoi diagram with the
sphere
\end{enumerate}
\end{theorem}
Note that although the divergence is not defined in the pure states, we
can consider the limit of the diagram with respect to the divergence in
the pure states. In this paper, Voronoi sites are plotted as a first
argument of the divergence $D(\cdot||\cdot)$.  From now on, when we say
``Voronoi diagram with respect to the divergence'', it means a limit of
a diagram with sites in the first argument of $D(\cdot||\cdot)$.

\Section{Higher level case}\label{sec-higher-dim}
In this section, we show that the coincidence which happens in one-qubit
case never occurs in a higher level case. To show it, it is enough to
look at some section of the diagrams with some
hyperplain. If the diagrams do not coincide in the section, you can say
they are different.

Suppose that $d\geq 3$ and that the space of general quantum states is
expressed as Equation~(\ref{multi-level-rho}), and let us think the
section of it with a hyperplain:
\begin{equation} \label{rep-hyperplain}
 \xi_{d+2} = \xi_{d+3} = \cdots = \xi_{d^2-1}.
\end{equation}
Then the section is expressed as:
\begin{eqnarray} \label{section-rho}
\rho=\!\!
\left(
\begin{array}{ccccc}
\frac{\xi_1+1}{d} & \!\!\frac{\xi_{d}-i\xi_{d+1}}{2}& & & \smash{\lower1.0ex\hbox{\bg 0}}\\
\frac{\xi_{d}+i\xi_{d+1}}{2} & \!\!\frac{\xi_2+1}{d}& & & \\
 & & \!\!\!\!\!\!\!\!\ddots & & \\
 & & & \!\!\!\!\!\!\!\!\frac{\xi_{d-1}+1}{d}&\\
\smash{\hbox{\bg 0}} & & & &
 \!\!\!\!\!\!\!\!\frac{-\sum_{i=1}^{d-1}\xi_i+1}{d}
\end{array}
\right)\!
.\!
\end{eqnarray}
The elements of this matrix are 0 except diagonal, (0,1), and (1,0)
elements. This matrix is diagonalized with a unitary matrix as:
\begin{eqnarray}\label{diagonalized}
\lefteqn{\rho=
\left(
\begin{array}{cc}
 X & 0 \\
 0 & I_{d-2}
\end{array}
\right)}
\nonumber\\
&\times
\left(
\begin{array}{cccccc}
\lambda_1 & & & & & \\
 & \lambda_2 & & & & \\
 & &\frac{\xi_3+1}{d} & & & \\
 & & &\ddots & & \\
 & & & &\frac{\xi_{d-1}+1}{d} & \\
 & & & & &\frac{-\sum_{j=1}^{d-1} \xi_j + 1}{d}
\end{array}
\right)\nonumber\\
&\times
\left(
\begin{array}{cc}
 X^* & 0 \\
 0 & I_{d-2}
\end{array}
\right),
\end{eqnarray}
where
\begin{eqnarray}
r = \sqrt{\frac{(\xi_1-\xi_2)^2}{d^2}+\xi_d^2+\xi_{d+1}^2},
\end{eqnarray}
\begin{eqnarray}
\lambda_1=\frac{\xi_1+\xi_2+2}{2d}+\frac{r}{2}, \\
\lambda_2=\frac{\xi_1+\xi_2+2}{2d}-\frac{r}{2},
\end{eqnarray}

\begin{eqnarray}
X=
\left(
{
\begin{array}{cc}
\displaystyle\frac{\frac{\xi_d - i\xi_{d+1}}{2}}
{\sqrt{R_+}}
 &
\displaystyle\frac{\frac{\xi_d - i\xi_{d+1}}{2}}
{\sqrt{R_-}}
\vspace{3mm}
\\
\displaystyle \frac{\frac{\xi_2-\xi_1}{2d}+\frac{r}{2}}
{\sqrt{R_+}}
&
\displaystyle \frac{\frac{\xi_2-\xi_1}{2d}-\frac{r}{2}}
{\sqrt{R_-}}
\end{array}
}
\right),\\
 R_+ =
  \frac{\xi_d^2+\xi_{d+1}^2}{4}+\left(\frac{\xi_2-\xi_1}{2d}+\frac{r}{2}\right)^2,\\
 R_- =
  \frac{\xi_d^2+\xi_{d+1}^2}{4}+\left(\frac{\xi_2-\xi_1}{2d}-\frac{r}{2}\right)^2.
\end{eqnarray}

Now we will figure out the necessary and sufficient condition for the diagonal
matrix of Equation~(\ref{diagonalized}) to be rank $1$. For that condition to
hold, the following three cases can be considered:
\paragraph{Case 1}
(only $d$-th raw of the matrix is non-zero)
\begin{equation}
 \xi_1=\xi_2=\cdots=\xi_{d-1}=-1,\, \xi_d=\xi_{d+1}=0.\nonumber
\end{equation}
\paragraph{Case 2}
 (only one $i$-th raw ($3\leq i \leq d-1$) is non-zero)
\begin{equation}
 \xi_1=\xi_2=-1,\xi_d=\xi_{d+1}=0,\nonumber
\end{equation}
all of $\xi_j\,(3\leq j \leq d-1)$ are $-1$ except one (let its index to be $k$) and 
$\xi_k=d-3$.

\paragraph{Case 3}
(only $\lambda_2$ is non-zero)
\begin{multline}
 \xi_1+\xi_2=d-2,\,
\frac{\xi_2- \xi_1}{d^2} +\frac{d^2}{4}(\xi_d^2+\xi_{d+1}^2) = 1, \\
  \xi_3=\xi_4=\cdots=\xi_{d-1}=-1.
\end{multline}
Note that it is impossible that only $\lambda_1$ is non-zero.
In both Case~1 and Case~2, the set of points that satisfies the condition
is just one point, so our main interest is Case~3. The
set of points that satisfies this condition is a manifold. Actually,
Case~3 satisfies
\begin{equation}\label{ellipsoid-pure}
\frac{(d-2- 2\xi_1)^2}{d^2} +\frac{d^2}{4}(\xi_d^2+\xi_{d+1}^2) = 1, \,
\end{equation}
and this is an ellipsoid.

Then we prepare for workout of the divergence.
The log of $\rho$
is expressed as:
\begin{eqnarray}
\lefteqn{ \log \rho=}\nonumber\\
&
\left(
\begin{array}{cc}
 X & 0 \\
 0 & I_{d-2}
\end{array}
\right)\nonumber\\
&\!\!\!\!\times\!\!
\left(
\begin{array}{cccccc}
\log\lambda_1 & & & & & \\
 & \!\!\!\!\!\!\!\!\!\log\lambda_2 & & & & \\
 & &\!\!\!\!\!\!\!\!\!\log\frac{\xi_3+1}{d} & & & \\
 & & &\ddots & & \\
 & & & &\!\!\!\!\!\!\!\!\!\log\frac{\xi_{d-1}+1}{d} & \\
 & & & & &\!\!\!\!\!\!\!\!\!\log\frac{-\sum_{j=1}^{d-1} \xi_j + 1}{d}
\end{array}
\right)\!\!\!\!\!\!\!\nonumber\\
&\times
\left(
\begin{array}{cc}
 X^* & 0 \\
 0 & I_{d-2}
\end{array}
\right).
\end{eqnarray}
Thus, we obtain
\begin{eqnarray} 
\lefteqn{\Tr \sigma \log \rho =
 \frac{\eta_1+1}{d}\cdot\frac{\xi_d^2+\xi_{d+1}^2}{4}\left[\frac{\log
 \lambda_1}{R_+}+\frac{\log \lambda_2}{R_-}\right]}\nonumber\\
&+\frac{\eta_d\xi_d + \eta_{d+1}\xi_{d+1}}{2}\!\!
\left[
\frac{\frac{\xi_2-\xi_1}{2d}+\frac{r}{2}}{R_+}\!\log \lambda_1 +
\frac{\frac{\xi_2-\xi_1}{2d}-\frac{r}{2}}{R_-}\!\log \lambda_2 
\right]\nonumber\\
&+ \frac{\eta_2+1}{d}\left[
\frac{\left(\frac{\xi_2-\xi_1}{2d}+\frac{r}{2}\right)^2}{R_+}\log \lambda_1 +
\frac{\left(\frac{\xi_2-\xi_1}{2d}-\frac{r}{2}\right)^2}{R_-}\log \lambda_2
\right]\nonumber\\
&+ \frac{1-\xi_1-\xi_2}{d}.
\end{eqnarray}

With some workout, we get
\begin{eqnarray}
 R_+ = r\left(\frac{\xi_2-\xi_1}{2d}+\frac{r}{2}\right),
 R_- = -r\left(\frac{\xi_2-\xi_1}{2d}-\frac{r}{2}\right)\!\!.\!\!
\end{eqnarray}
Using these fact and the assumption $\eta_1+\eta_2=\xi_1+\xi_2=d-2$, we get
\begin{eqnarray}\label{simplified-rho}
\lefteqn{ \Tr\sigma\log\rho=}\nonumber\\
&\left[
\frac{\eta_d\xi_d+\eta_{d+1}\xi_{d+1}}{2r}
+\frac{2\left(\eta_1-\frac{d-2}{2}\right)\left(\xi_1-\frac{d-2}{2}\right)}{d^2r}
\right]\log \frac{\lambda_1}{\lambda_2}\nonumber\\
&+\frac{1}{2}\log \lambda_1 \lambda_2.
\end{eqnarray}

Next we think of a Voronoi diagram with only two regions for
simplicity. It is enough for our objective. Let $\sigma$ and
$\tilde{\sigma}$ be two sites, and suppose that $\rho$ moves along the
boundary of the Voronoi regions.  Suppose that $\sigma$ and
$\tilde{\sigma}$ are parameterized by $\{\eta_j\}$ and
$\{\tilde{\eta}_j\}$ respectively in the same way as $\rho$.

We consider what happens if $r(0\leq r<1)$ is fixed and the following holds:
\begin{eqnarray}\label{ellipsoid-shrink}
 \xi_1+\xi_2=d-2,\, \xi_3=\cdots=\xi_{d-1}=-1.
\end{eqnarray}
The condition $0\leq r <1$ means that $\rho$ is semi-positive and not a
pure state while $r=1$ in pure states.  In other words, we regard that
$\rho$ is on the same ellipsoid obtained by shrinking the ellipsoid
expressed by Equation~(\ref{ellipsoid-pure}). These settings are in
order to take a limit of a diagram to get a diagram in the pure
states. Taking the limit $r\to 1$, we can get a condition for pure
states. This procedure is analogous to the method used in \cite{kato05}.

Now to think of the shape of boundary, we have to solve the equation
\begin{eqnarray}
D(\sigma||\rho)=D(\tilde{\sigma}||\rho),
\end{eqnarray}
and this is equivalent to
\begin{eqnarray}
 \Tr (\sigma - \tilde{\sigma})\log\rho = 0.
\end{eqnarray}
Using Equation~(\ref{simplified-rho}), we obtain
\begin{eqnarray}\label{boundary-result}
\lefteqn{  \Tr (\sigma - \tilde{\sigma})\log\rho =}\nonumber\\
&\!\!\!\!\!\!\frac{1}{2r}
\left[{\footnotesize
(\eta_d\!-\!\tilde{\eta}_d)\xi_d + (\eta_{d+1} \!-\! \tilde{\eta}_{d+1})\xi_{d+1}
+\frac{4(\eta_1\!-\!\tilde{\eta}_1)\left(\xi_1\!-\!\frac{d-2}{2}\right)}{d^2}}
\right]
\!\!\!\!\!\!\!\!\!\!\!\!\nonumber\\
&\times\log \frac{\lambda_1}{\lambda_2}.
\end{eqnarray}
Here when $r=0$, this is zero because $\lambda_1/\lambda_2=1$. In that
case, $\rho$ can take only one point, but we do not have to care about
this case because we are going to take the limit $r\to1$. From now on,
we suppose $r>0$ and that means $\lambda_1/\lambda_2\neq 1$.

Hence we get the following equation that holds in the boundary of the
Voronoi diagram:
\begin{eqnarray}\label{boundary-divergence}
(\eta_d-\tilde{\eta}_d)\xi_d + (\eta_{d+1} - \tilde{\eta}_{d+1})\xi_{d+1}
\nonumber\\
+\frac{4(\eta_1-\tilde{\eta}_1)\left(\xi_1-\frac{d-2}{2}\right)}{d^2}
=0.
\end{eqnarray}
Consequently, taking the limit $r\to 1$, we get
Equation~(\ref{boundary-divergence}) as the expression of the boundary in
pure states.

A careful inspection of Equation~(\ref{boundary-divergence}) tells us a
geometric interpretation of this boundary. We obtain the following
theorem:
\begin{theorem}
 On the ellipsoid of the pure states which appears in the section with
 the hyperplain defined above, if transfered by a linear transform which
 maps the ellipsoid to a sphere, the Voronoi diagram with respect to the
 divergence coincides with the one with respect to the geodesic
 distance.
\end{theorem}
\begin{proof}
 Think of the affine transform defined by
\begin{eqnarray}
 \left(
\begin{array}{c}
 x\\
 y\\
 z
\end{array}
\right)
=
 \left(
\begin{array}{c}
 \frac{\xi_1-\frac{d-2}{2}}{\frac{d}{2}}\\
 \xi_d\\
 \xi_{d+1}
\end{array}
\right),
\end{eqnarray}
then Equation~(\ref{boundary-divergence}) is expressed as
\begin{eqnarray}
 x'(x-\tilde{x})+y'(y-\tilde{y})+z'(z-\tilde{z})=0,
\end{eqnarray}
while Equation~(\ref{ellipsoid-pure}) becomes
\begin{eqnarray}
 x^2+y^2+z^2=1.
\end{eqnarray}
Thus when $(x,y,z)$ and $(\tilde{x},\tilde{y},\tilde{z})$ are fixed, the
 point $(x',y',z')$ which stand for $\eta$ runs along the geodesic.
\end{proof}

Now we work out the Voronoi diagram with respect to Euclidean
distance. Under the assumption above, the Euclidean distance is
expressed as
\begin{eqnarray}
\lefteqn{ d(\sigma,\rho)}\nonumber\\
&\!\!\!\!=\!(\eta_1\!-\!\xi_1)^2\!+\!(\eta_2\!-\!\xi_2)^2\!+\!
(\eta_d\!-\!\xi_d)^2 \!+\! (\eta_{d+1}\!-\!\xi_{d+1})^2\!\!\!\!\nonumber\\
&=2(\eta_1-\xi_1)^2+
(\eta_d-\xi_d)^2 + (\eta_{d+1}-\xi_{d+1})^2,
\end{eqnarray}
and we get the equation for boundary as
\begin{multline}\label{boundary-euclidean}
 d(\sigma,\rho)-d(\tilde{\sigma},\rho)=\\
-4(\eta_1-\tilde{\eta}_1)\xi_1
-2(\eta_d-\tilde{\eta}_d)\xi_d
-2(\eta_{d+1}-\tilde{\eta}_{d+1})\xi_{d+1}\\
+2(\eta_1^2-\tilde{\eta}_1^2)+(\eta_d^2-\tilde{\eta}_d^2)
+(\eta_{d+1}^2-\tilde{\eta}_{d+1}^2)=0.
\end{multline}
By comparing the coefficients of $\xi_1$, $\xi_d$, and $\xi_{d+1}$,
we can tell that the boundaries expressed by
Equation~(\ref{boundary-divergence}) and (\ref{boundary-euclidean}) are
different. To show how different they are, we give some examples in the
rest of this section.

\begin{figure}[!h]
\begin{center}
 \includegraphics[scale=.3,clip]{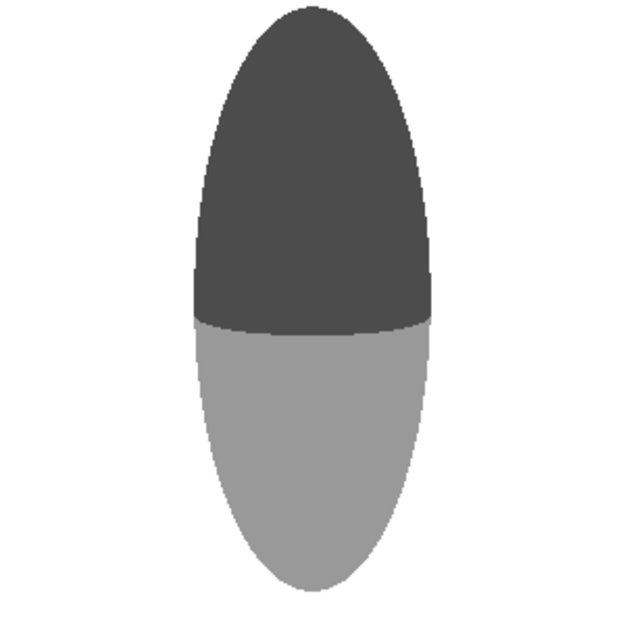}\hspace{-1cm}
 \includegraphics[scale=.3,clip]{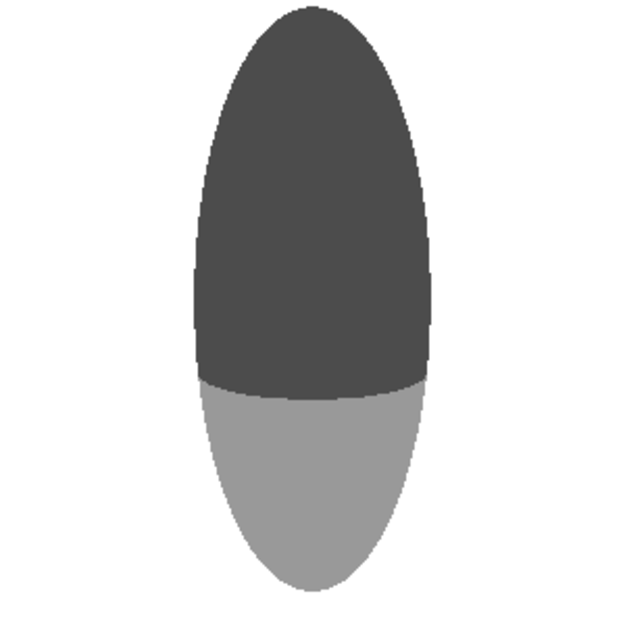} \caption{An example of
 a Voronoi diagram with two sites. The figure on the left is the diagram
 by the divergence, and the figure on the right is the diagram by the
 Euclidean distance.}  \label{fig-example1}
\end{center}
\end{figure}

\begin{example}
 Suppose that $(\eta_1,\eta_d,\eta_{d+1})=(d-1,0,0)$ and
$(\tilde{\eta_1},\tilde{\eta}_d,\tilde{\eta}_{d+1})=(-1,0,0)$, then the
boundary is a) $\xi_1=\frac{d-2}{2}$ for the divergence, and b)
$\xi_1=1$ for the Euclidean distance.  Fig.\,\ref{fig-example1} shows
this example for $d=5$. In Fig.\,\ref{fig-example1}, Voronoi sites are
located on the top and the bottom of the ellipsoid.  The two diagrams
are the same when $d=4$, but are different otherwise.
\end{example}

\begin{example}
 Suppose that $(\eta_1,\eta_d,\eta_{d+1})=(0,1,0)$ and
$(\tilde{\eta_1},\tilde{\eta}_d,\tilde{\eta}_{d+1})=(0,-1,0)$, then
the boundary is, for both the divergence and Euclidean distance,
expressed by $\xi_{d+1}=0$.
\end{example}

\begin{figure}[!h]
\begin{center}
 \includegraphics[scale=.3,clip]{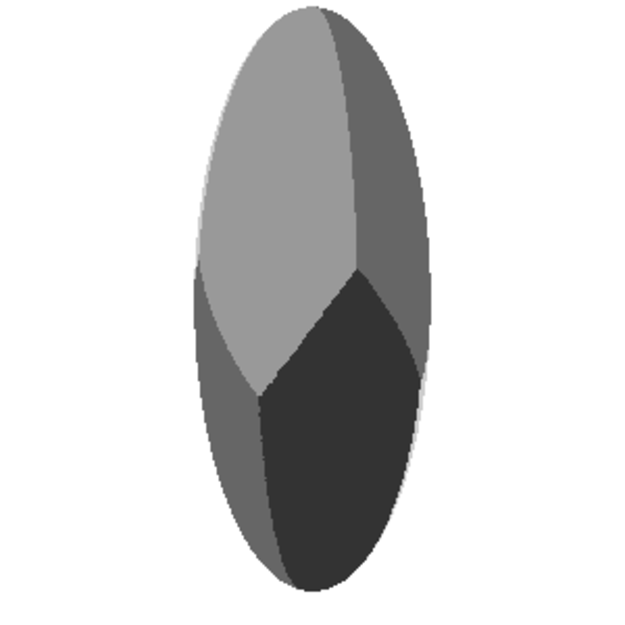}\hspace{-1cm}
 \includegraphics[scale=.3,clip]{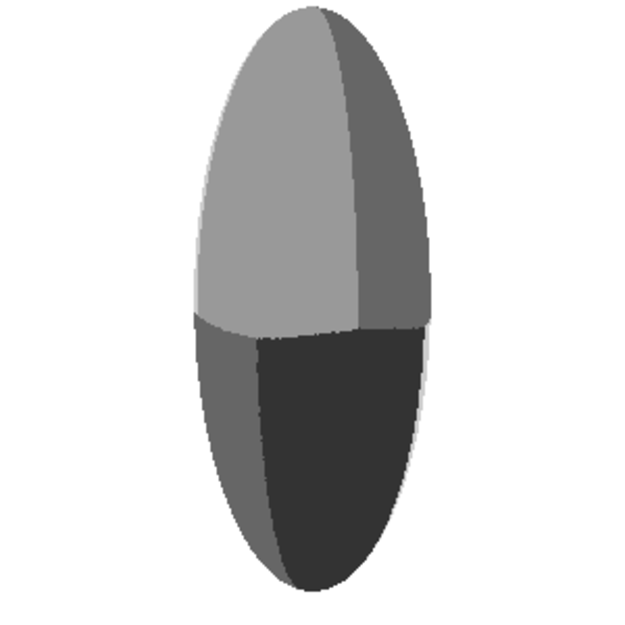} \caption{An example of
 a Voronoi diagram with eight sites. The left is the diagram by the
 divergence, and the right is by the Euclidean distance.}
 \label{fig-example3}
\end{center}
\end{figure}

\begin{example}
Consider the Voronoi diagram with the following eight sites:
\begin{eqnarray}
 \left(\frac{d-2}{2}+\frac{d}{2\sqrt{3}},\,\pm\frac{1}{\sqrt{3}},\,\pm\frac{1}{\sqrt{3}}\right),\nonumber\\
  \left(\frac{d-2}{2}-\frac{d}{2\sqrt{3}},\,\pm\sqrt{\frac{2}{3}},\,0\right),\nonumber\\
  \left(\frac{d-2}{2}-\frac{d}{2\sqrt{3}},\,0,\,\pm\sqrt{\frac{2}{3}}\right),
\end{eqnarray}
where $\pm$'s mean all the possible combinations. Then the Voronoi diagrams
 look like Fig.\,\ref{fig-example3}. This figure is also for
 $d=5$. Obviously they are different.
\end{example}

\Section{Conclusion}\label{sec-conclusion}

We proved that in $n$-level system for $n \geq 3$, the Voronoi diagrams
with respect to the divergence and Euclidean distance do not
coincide. Additionally we obtained an explicit expression of some section
of the boundary
of the Voronoi diagram with respect to the divergence. The
section is an ellipsoid and after some linear transform, the boundary
becomes a geodesic on a sphere. Interestingly this is similar to the
whole space of the
one-qubit states even though a space of higher level has much more
complicated geometric structure.

We also showed some geometric structure concerning how pure states appear in a
whole quantum states. Although our result is very restricted, we believe
this will be a help for further understanding of the structure. To relax
the restriction is our future work. 

The result shown in this paper depends on the parameterization of
density matrix. The parameterization used in this paper, though it is
very natural one, is not unique. To think of another parameterization is
another future work.

\paragraph{Acknowledgment}
We would like to thank Dr.~Masahito Hayashi who pointed out a serious
logical gap in the earlier version of this paper.

\bibliography{ISVD2006-kato}
\bibliographystyle{latex8}

\end{document}